# Real-space identification of distinct magnetic configurations in a candidate *d*-wave altermagnet

Jin-Cheng Gu[1,2,#], Mingzhe Hu[1,3,#], Ziyin Song[1,3,#], Lihan Wang[1,3,#], Lihong Wang[1,3], Junming Zhang[1,3], Jiali Zhao[1], Hang Li[1], Shifeng Jin[1], Xin-Ding Zhang[2,†], Genfu Chen[1,3,†], Hongming Weng[1,3,†], Zhongxu Wei[1,†], Tian Qian[1,†]

[1]Beijing National Laboratory for Condensed Matter Physics, Institute of Physics, Chinese Academy of Sciences, Beijing 100190, China.

[2]Guangdong Provincial Key Laboratory of Quantum Engineering and Quantum Materials, Guangdong-Hong Kong Joint Laboratory of Quantum Matter, and Frontier Research Institute for Physics, South China Normal University, Guangzhou 510006, China.

[3]University of Chinese Academy of Sciences, Beijing 100490, China

#These authors contributed equally to this work.

†Corresponding authors: xdzhang@scnu.edu.cn; gfchen@iphy.ac.cn; hmweng@iphy.ac.cn; zhongxuwei@iphy.ac.cn; tqian@iphy.ac.cn

**Altermagnetism is an emerging class of magnetic order characterized by momentum-dependent spin-split electronic structures despite vanishing net magnetization. Although momentum-space signatures consistent with altermagnetism have been reported in a growing number of materials, their relationship to the underlying real-space magnetic configurations remains incompletely understood, because similar spin-split electronic structures can arise from distinct magnetic orders. In the candidate *d*-wave altermagnet $KV_2Se_2O$, the magnetic origin of the observed momentum-dependent spin splitting has remained controversial. Here, we employ spin-polarized scanning tunnelling microscopy combined with magnetic-field-dependent quasiparticle interference imaging to determine the magnetic configuration of $KV_2Se_2O$ at the atomic scale. Spin-resolved quasiparticle interference reveals a checkerboard-like antiparallel spin texture within the $V_2O$ layer and determines its interlayer spin arrangement across unit-cell step edges. Remarkably, we identify both C-type and G-type magnetic configurations, both of which generate similar spin-split electronic structures at the single-layer level but correspond to *d*-wave altermagnetic and conventional antiferromagnetic orders, respectively. These observations reveal a complex magnetic landscape arising from nearly degenerate magnetic states. Our results establish a direct connection between momentum-space spin splitting and real-space magnetic order, providing a framework for identifying the microscopic origin of spin-split electronic structures in altermagnetic materials.**

Altermagnetism has recently emerged as a distinct class of magnetic order that combines compensated magnetism, characteristic of antiferromagnets, with momentum-dependent spin-split electronic structures typically associated with ferromagnets[1–6]. Unlike conventional antiferromagnets, in which opposite-spin sublattices are related by translation or inversion symmetry, altermagnets are governed by crystal rotation, mirror, glide, or screw symmetries, enabling anisotropic spin splitting even in the absence of spin–orbit coupling[7–16]. The coexistence of vanishing net magnetization and spin-split electronic structures has attracted considerable attention owing to its potential implications for spin transport, correlated quantum phases, and topological phenomena[17–28].

A major experimental advance in this field has been the observation of momentum-dependent spin splitting by spin- and angle-resolved photoemission spectroscopy (SARPES), which has established key momentum-space signatures of altermagnetism in several candidate materials[12,13,15,16,29–33]. However, momentum-space spin splitting alone is insufficient to uniquely identify altermagnetic order. Certain conventional antiferromagnetic configurations can also generate local spin-split electronic structures, giving rise to similar spectroscopic signatures. Establishing the microscopic origin of these spin-split electronic structures therefore requires direct information about the underlying real-space magnetic configurations. Spin-polarized scanning tunnelling microscopy (SP-STM) provides an atomic-scale probe of magnetic structure in real space and can resolve its evolution across neighbouring atomic layers at the surface[34–38], offering a direct route to connect spin-split electronic structures with their underlying magnetic configurations.

The layered vanadium oxychalcogenides $AV_2Q_2O$ (A = K, Rb, Cs; $Q$ = Se, Te) have emerged as prominent candidates for metallic room-temperature altermagnetism[15,16,39,40]. SARPES measurements have revealed $C_2$-symmetric momentum-dependent spin splitting, consistent with $d$-wave altermagnetic order[15,16]. These observations are well reproduced by first-principles calculations based on a checkerboard-like antiparallel spin texture of the V sublattices within each $V_2O$ layer and a C-type interlayer stacking configuration[15,16]. However, subsequent neutron diffraction measurements identified a G-type magnetic structure, corresponding to a conventional antiferromagnetic state rather than an altermagnetic state[40,41]. Both C-type and G-type magnetic structures share the same checkerboard spin texture within an individual $V_2O$ layer and therefore generate similar spin-split electronic structures at the single-layer level (Extended Data Fig. 1), which could give rise to similar ARPES signatures. Yet their distinct interlayer symmetry relations lead to fundamentally different global electronic states. This exposes a challenge in establishing altermagnetic order from momentum-space signatures alone. Determining the real-space magnetic configuration is therefore essential for establishing the microscopic magnetic origin of the observed spin-split electronic structure.

In this work, we investigate the magnetic configuration of $KV_2Se_2O$ at the atomic scale using SP-STM combined with magnetic-field-dependent quasiparticle interference (QPI) imaging. By resolving the spin texture within individual $V_2O$ layers and identifying the interlayer spin arrangement across unit-cell step edges, we directly determine the magnetic configuration underlying the observed spin-split electronic structure. These measurements provide a real-space framework for connecting momentum-space signatures of altermagnetism with their microscopic magnetic origin.

## Spin-dependent QPI framework

Figure 1 illustrates the spin-dependent QPI expected for C-type and G-type magnetic structures in $KV_2Se_2O$ and outlines the experimental strategy used to determine the real-space magnetic configuration. The crystal structure consists of stacked $V_2O$ square lattices separated by K and Se layers[42–44]. The magnetic moments on the V sites are predominantly oriented along the $c$ axis, as indicated by previous nuclear magnetic resonance and neutron diffraction measurements[15,41], and form a checkerboard-like antiparallel spin texture within each $V_2O$ layer (Fig. 1a,b). This spin texture gives rise to an electronic structure comprising two pairs of quasi-one-dimensional (1D) Fermi surface (FS) sheets extending along $k_x$ and $k_y$, respectively, together with electron-like FS pockets at the Brillouin zone boundary.

The interlayer evolution of the spin texture leads to distinct FS topologies. In the C-type magnetic structure, the spin texture remains identical upon stacking along the $c$ axis, resulting in identical spin-polarized FSs in adjacent $V_2O$ layers and globally spin-polarized FSs (Fig. 1a). In contrast, in the G-type magnetic structure, the spin texture reverses between adjacent layers, giving rise to oppositely spin-polarized FSs in adjacent $V_2O$ layers and globally spin-degenerate FSs (Fig. 1b). First-principles calculations show that the interlayer exchange coupling $J$ remains negative near the experimental $c$-axis lattice constant of $KV_2Se_2O$, indicating that the G-type magnetic configuration is energetically favoured (Fig. 1d). However, its small magnitude (~1 meV) implies only a weak energetic preference over the C-type configuration.

Because $KV_2Se_2O$ cleaves preferentially within the K layer, the magnetic $V_2O$ layer is not directly accessible to SP-STM. To probe the in-plane spin texture, we employ spin-resolved QPI imaging. Interband scattering between the quasi-1D FS sheets is expected to generate anisotropic QPI modulations. Because the two FS pairs carry opposite spin polarizations, the corresponding scattering channels are associated with opposite spin character (Fig. 1c). Since the tunnelling probability depends on the relative spin orientation between the magnetic tip and the spin-polarized electronic states, the QPI intensity is sensitive to the tip magnetization. Reversing the field, and thus the tip magnetization, modifies the tunnelling probability into spin-polarized electronic states and produces opposite intensity changes in the two scattering channels (Fig. 1e). Combined with the correspondence between the scattering channels and the FS sheets, the field-dependent QPI measurements enable determination of the spin polarization of the electronic structure and thereby the underlying real-space spin texture.

The interlayer evolution of the spin texture can be examined by extending the same analysis across a unit-cell step edge (Fig. 1f). In the C-type configuration, the real-space spin texture remains identical between adjacent layers, and the field-dependent QPI contrast is therefore expected to remain unchanged across the step edge. In contrast, the checkerboard spin texture reverses between adjacent layers in the G-type configuration, leading to a reversal of the field-dependent QPI contrast across the step edge. Spin-polarized STM measurements across unit-cell step edges therefore provide a direct real-space probe of the layer-dependent magnetic configuration at the surface of $KV_2Se_2O$.

**Identification of QPI scattering channels**

$KV_2Se_2O$ cleaves within the K layer, producing a half-occupied surface that reconstructs into a $\sqrt{2} \times \sqrt{2}$ lattice, as confirmed by atomically resolved STM measurements (Fig. 2a). The pronounced dark feature in the topography corresponds to an isolated K vacancy. Figure 2b compares differential conductance spectra acquired far from and near a K vacancy. Away from the defect, the spectrum exhibits a ~20 meV gap at the Fermi level, previously attributed to a hybridization gap associated with FS nesting induced by an additional spin-density-wave order[15]. In contrast, pronounced in-gap states emerge near the vacancy, indicating that the defect strongly perturbs the local low-energy electronic states and gives rise to bound states.

K vacancies act as scattering centres, giving rise to low-energy QPI that is dominated by these bound states. Figure 2c shows a differential conductance map $g(r)$ acquired at 6 meV, revealing standing-wave patterns arising from QPI along the $a$ and $b$ axes. A representative line profile extracted from the conductance map yields a modulation period of ~0.87 nm (Fig. 2d). The Fourier-transformed map $g(q)$ resolves two dominant scattering vectors, $q_1$ and $q_2$, oriented along $q_a$ and $q_b$, respectively (Fig. 2e). Both vectors have magnitudes of ~7.76 $nm^{-1}$, corresponding to a real-space period of ~0.81 nm, consistent with the line-profile analysis. To identify the underlying scattering processes, we compare the QPI wave vectors with the FS topology measured by ARPES (Fig. 2g). The $q_1$ ($q_2$) vector matches the momentum separation between the quasi-1D $\alpha_1$–$\alpha_2$ ($\beta_1$–$\beta_2$) FS sheets, indicating that the QPI originates from interband scattering between these sheets. This assignment is further supported by the energy dependence of $q_1$, which closely tracks the $\alpha_2$ band dispersion (Fig. 2f).

As a reference for spin-polarized measurements, we first examine the magnetic-field dependence of the QPI using a non-magnetic tip. Differential conductance maps acquired under opposite out-of-plane magnetic fields of ±8 T, labelled $g_\uparrow(r)$ and $g_\downarrow(r)$, are nearly identical (Fig. 2h,i). Accordingly, the difference map $g_\uparrow(r) - g_\downarrow(r)$ is featureless within experimental resolution (Fig. 2j), indicating that magnetic-field reversal does not measurably affect the QPI signal in the absence of spin-selective tunnelling. This establishes a baseline for subsequent spin-polarized measurements, where field-dependent QPI contrast can be attributed to spin-dependent tunnelling.

**In-plane spin texture from QPI**

We next use spin-resolved QPI imaging to determine the spin texture within an individual $V_2O$ layer. Figure 3a,b shows differential conductance maps $g_\uparrow(r)$ and $g_\downarrow(r)$ acquired with a magnetic Cr tip under opposite out-of-plane magnetic fields of ±8 T. Both datasets exhibit anisotropic QPI modulations consistent with those observed in spin-unpolarized measurements. Although the overall QPI patterns under opposite magnetic fields appear similar, the difference map $g_\uparrow(r) - g_\downarrow(r)$ reveals positive contrast along the $a$ axis and negative contrast along the $b$ axis, indicating a pronounced spin-dependent contribution to the QPI (Fig. 3c). Line profiles extracted along both directions quantitatively confirm the opposite responses of the QPI intensity upon magnetic-field reversal (Fig. 3d,e). The field-induced QPI intensity change is positive along the $a$ axis and negative along the $b$ axis, highlighting a pronounced anisotropic spin-dependent contrast (Fig. 3f).

The Fourier-transformed maps $g_\uparrow(q)$ and $g_\downarrow(q)$ (Fig. 3g,h) reveal two dominant scattering vectors $q_1$ and $q_2$, consistent with those identified in spin-unpolarized measurements. The difference map $g_\uparrow(q) - g_\downarrow(q)$ shows positive contrast at $q_1$ and negative contrast at $q_2$ (Fig. 3i). Similar spin-dependent QPI responses are observed in a different sample (Extended Data Fig. 2). The experimentally observed contrast distribution is well reproduced by the calculated spin-contrast QPI from a joint density of states (JDOS) simulation based on the spin-polarized FSs, with $q_1$ and $q_2$ associated with spin-up and spin-down scattering channels, respectively (Fig. 3j). The spin-dependent QPI response therefore identifies the $\alpha_{1,2}$ and $\beta_{1,2}$ FS sheets as spin-up and spin-down, respectively (Fig. 3k). The $\alpha_{1,2}$ and $\beta_{1,2}$ FSs originate from two V sublattices associated with V–O chains running along the *a* and *b* axes, respectively[15]. Accordingly, the V sublattice associated with the *a*-axis chains is assigned spin-up, while that associated with the *b*-axis chains is assigned spin-down, thereby establishing the checkerboard spin texture within an individual $V_2O$ layer (Fig. 3l).

**Distinct interlayer spin arrangements**

Having determined the spin texture within an individual $V_2O$ layer, we next investigate its evolution across adjacent layers. Figure 4a,g shows STM topographies of two representative unit-cell step edges on different $KV_2Se_2O$ surfaces. The measured step heights of 0.68 and 0.76 nm are close to the *c*-axis lattice constant of 0.73 nm[44], confirming that the upper and lower terraces correspond to adjacent $V_2O$ layers.

We first examine the step edge shown in Fig. 4a. Applying the same spin-resolved QPI analysis as in Fig. 3, we construct the difference maps $g_{\uparrow}(r) - g_{\downarrow}(r)$ on both terraces (Fig. 4c,d). The difference maps measured on the two terraces exhibit the same spin-dependent QPI response, with positive contrast along the *a* axis and negative contrast along the *b* axis. The corresponding Fourier-transformed difference maps $g_{\uparrow}(q) - g_{\downarrow}(q)$ also display the same sign distribution on both terraces, with positive contrast at $q_1$ and negative contrast at $q_2$ (Fig. 4e,f). This behaviour indicates that the spin texture remains unchanged across the unit-cell step edge and is therefore consistent with a C-type magnetic configuration (Fig. 4b).

In contrast, the step edge shown in Fig. 4g exhibits a qualitatively different behaviour. The real-space difference maps $g_{\uparrow}(r) - g_{\downarrow}(r)$ measured on the two terraces display opposite spin-dependent QPI responses (Fig. 4i,j). Correspondingly, the signs of the $q_1$ and $q_2$ features in the Fourier-transformed difference maps $g_{\uparrow}(q) - g_{\downarrow}(q)$ are reversed between the lower and upper terraces (Fig. 4k,l). Such a contrast reversal indicates that the spin texture reverses across the unit-cell step edge, consistent with a G-type magnetic configuration (Fig. 4h).

Taken together, these observations establish the presence of both C-type and G-type magnetic configurations at the surface of $KV_2Se_2O$. The coexistence is consistent with the weak interlayer exchange coupling and the resulting small energy difference between the two competing magnetic states (Fig. 1d).

**Summary and Outlook**

We have determined the real-space magnetic configurations underlying the spin-split electronic structure of $KV_2Se_2O$ using SP-STM combined with magnetic-field-dependent QPI imaging. Spin-resolved QPI measurements establish a checkerboard-like antiparallel spin texture within individual $V_2O$ layers, while layer-resolved measurements across unit-cell step edges directly reveal the interlayer spin arrangement. Remarkably, we identify both C-type and G-type magnetic configurations at the surface. As these two magnetic configurations share nearly identical spin-split electronic structures at the single-layer level and can therefore generate similar momentum-space signatures, momentum-dependent spin splitting alone is insufficient to uniquely determine the underlying magnetic order. Our results highlight the necessity of combining momentum-space and real-space probes when establishing the microscopic magnetic origin of spin-split electronic structures in candidate altermagnets.

The observation of both C-type and G-type magnetic configurations reveals a competition between distinct interlayer spin arrangements in $KV_2Se_2O$. Such competing magnetic configurations introduce an additional degree of freedom in layered magnetic materials, suggesting that subtle perturbations such as lattice distortion, strain, surface relaxation, or dimensional confinement may strongly influence the realized magnetic state. The ability to manipulate competing interlayer magnetic configurations may therefore provide a route to control magnetic symmetry and the associated spin-split electronic structures, opening new opportunities for engineering magnetic and electronic properties in layered altermagnetic materials.

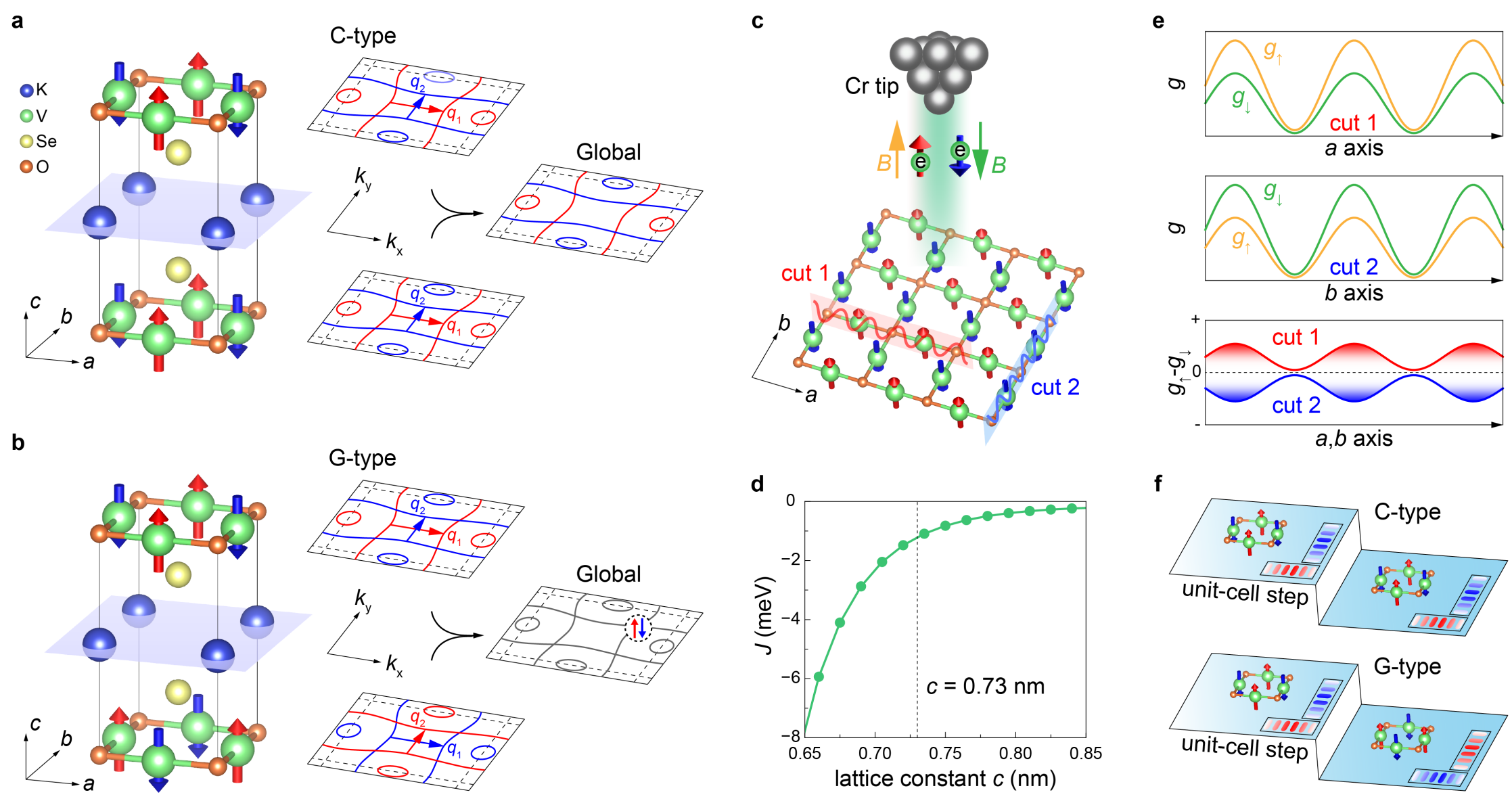

**Fig. 1 | Spin-dependent QPI expected for C-type and G-type magnetic structures. a**,**b**, Magnetic structures and corresponding FSs of $KV_2Se_2O$ in the C-type (**a**) and G-type (**b**) magnetic configurations. Red and blue arrows denote spin-up and spin-down moments on the V sites, respectively. The spin-polarized FSs of individual $V_2O$ layers are shown in the centre, and the resulting global FSs are shown on the right. $q_1$ and $q_2$ denote two dominant scattering channels. Red and blue denote spin-up and spin-down character, respectively. **c**, Schematic principle of spin-resolved QPI measurements using a Cr tip. The tip magnetization is controlled by an external magnetic field $B$ applied along the $c$ axis. Reversing the field modifies the tunnelling probability into spin-polarized electronic states, enabling spin-sensitive detection of QPI. Red and blue curves denote the scattering channels associated with spin-up and spin-down states along cut 1 (parallel to the $a$ axis) and cut 2 (parallel to the $b$ axis), respectively. **d**, Calculated interlayer exchange coupling $J$ as a function of the lattice constant $c$. Dashed vertical line marks the experimental $c$-axis lattice constant of $KV_2Se_2O$. **e**, Expected QPI intensity profiles $g_{\uparrow}(r)$ and $g_{\downarrow}(r)$ under opposite magnetic fields along cut 1 (upper) and cut 2 (middle). The corresponding difference $g_{\uparrow}(r) - g_{\downarrow}(r)$ is positive along cut 1 and negative along cut 2 (lower). **f**, Expected spin-dependent QPI contrast across a unit-cell step edge for the C-type (upper) and G-type (lower) magnetic structures. Red and blue denote positive and negative values of the QPI difference signal $g_{\uparrow}(r) - g_{\downarrow}(r)$, respectively.

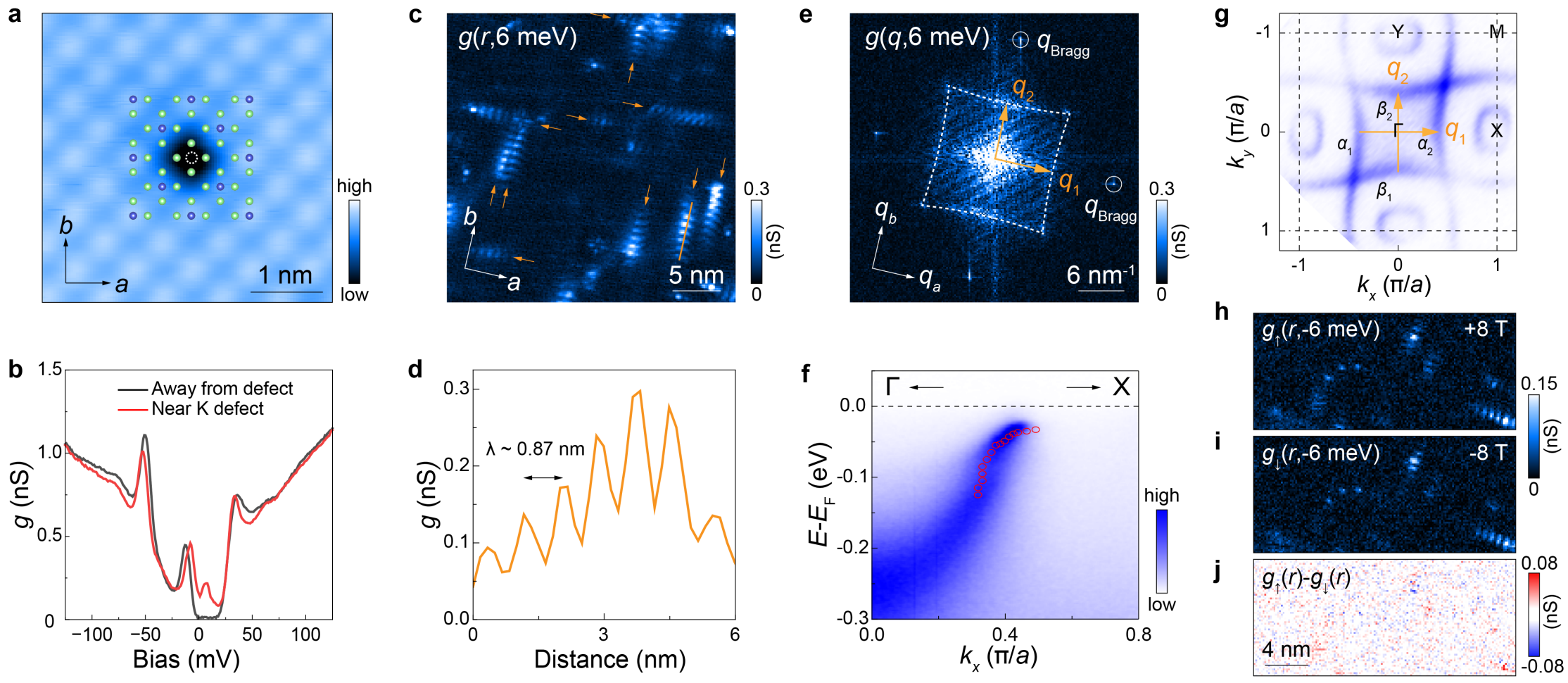

**Fig. 2 | QPI scattering channels in $KV_2Se_2O$. a**, Atomic-resolution STM topography ($V_s$ = 300 mV, $I_t$ = 300 pA) of the cleaved K layer. The dark contrast at the centre corresponds to a missing K atom. The overlaid structural model shows the positions of K and V atoms. **b**, Differential conductance spectra acquired away from (black) and near (red) a K vacancy. **c**, Differential conductance map $g(r)$ acquired at 6 meV. Yellow arrows indicate QPI standing-wave patterns. **d**, Line profile extracted along the yellow line in **c**, showing a modulation period of ~0.87 nm. **e**, Fourier-transformed map $g(q)$ obtained from **c**. $q_{\mathrm{Bragg}}$ corresponds to the reciprocal lattice vectors of the $V_2O$ layer, while $q_1$ and $q_2$ denote the dominant QPI wave vectors. **f**, Band dispersion along Γ–X obtained by ARPES. Red circles denote the dispersion extracted from QPI measurements. The QPI data are shifted upward by 15 meV for comparison. **g**, FS intensity map obtained by ARPES. The quasi-1D sheets are labelled $\alpha_1$, $\alpha_2$, $\beta_1$, and $\beta_2$. $q_1$ and $q_2$ denote the scattering vectors connecting the $\alpha_1$–$\alpha_2$ and $\beta_1$–$\beta_2$ sheets, respectively. **h**,**i**, Differential conductance maps $g_\uparrow(r)$ and $g_\downarrow(r)$ acquired at –6 meV under out-of-plane magnetic fields of +8 T (**h**) and –8 T (**i**) using a nonmagnetic tip. **j**, Difference map $g_\uparrow(r) - g_\downarrow(r)$. Red and blue denote positive and negative contrast, respectively.

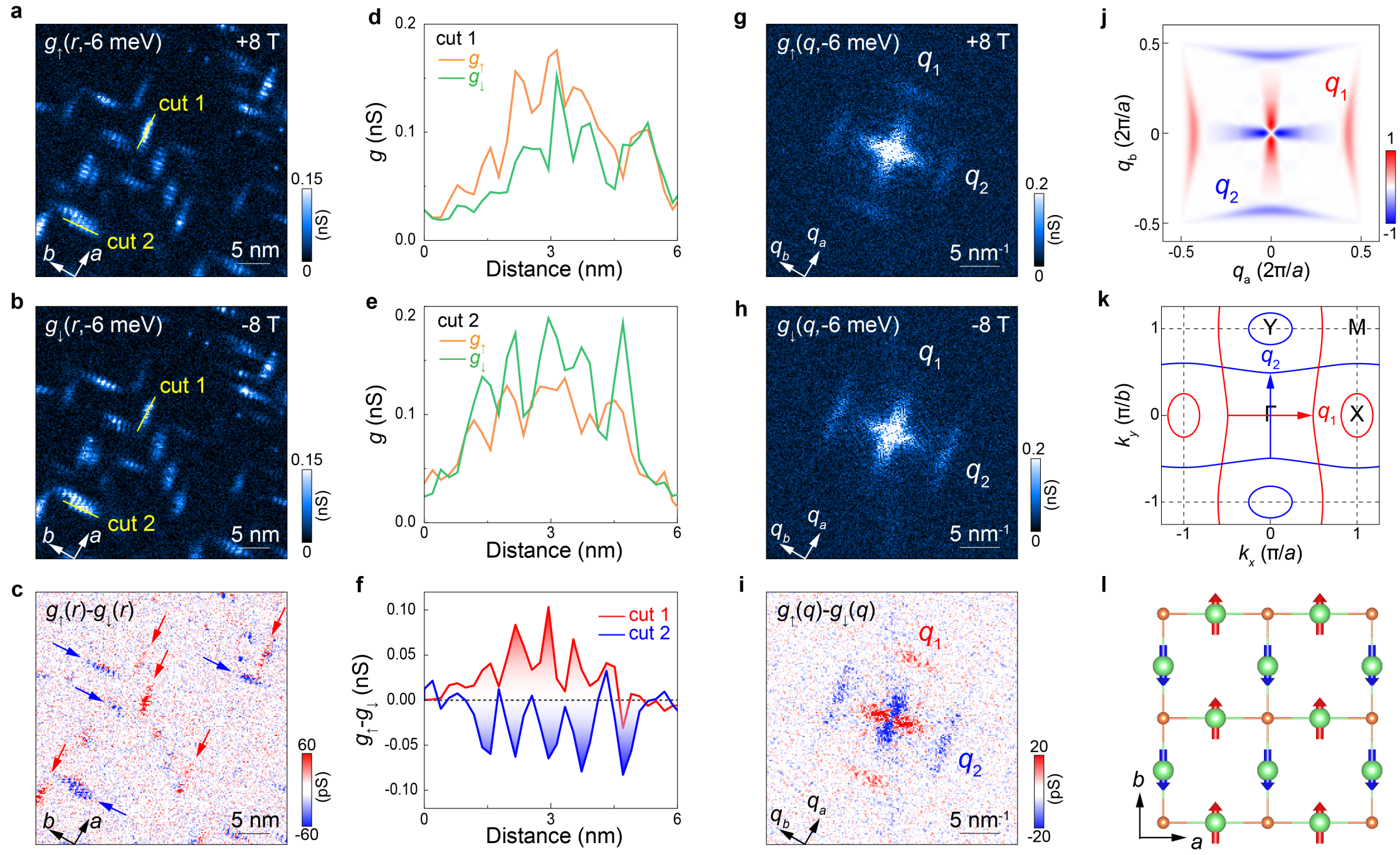

**Fig. 3 | Spin-resolved QPI in a $V_2O$ layer. a**,**b**, Differential conductance maps $g_{\uparrow}(r)$ and $g_{\downarrow}(r)$ acquired at −6 meV under out-of-plane magnetic fields of +8 T (**a**), and −8 T (**b**) using a magnetic Cr tip. Yellow lines indicate cut 1 and cut 2, which are parallel to the $a$ and $b$ axes, respectively. **c**, Difference map $g_{\uparrow}(r)-g_{\downarrow}(r)$. Red and blue arrows indicate positive and negative contrast, respectively. **d**,**e**, Line profiles of $g_{\uparrow}(r)$ and $g_{\downarrow}(r)$ extracted along cut 1 (**d**) and cut 2 (**e**). **f**, Line-profile difference $g_{\uparrow}(r)-g_{\downarrow}(r)$ along cut 1 and cut 2. **g**,**h**, Fourier-transformed maps $g_{\uparrow}(q)$ (**g**) and $g_{\downarrow}(q)$ (**h**) obtained from **a** and **b**, respectively. **i**, Fourier-transformed difference map $g_{\uparrow}(q)-g_{\downarrow}(q)$. **j**, Calculated spin-contrast QPI from a JDOS simulation at the Fermi level. The contrast is normalized to the range [−1,1]. **k**, Schematic FSs extracted from the ARPES data in Fig. 2g. Red and blue denote spin-up and spin-down character, respectively. **l**, $V_2O$ lattice with the corresponding spin configuration. Red and blue arrows denote spin-up and spin-down moments on the V sites, respectively.

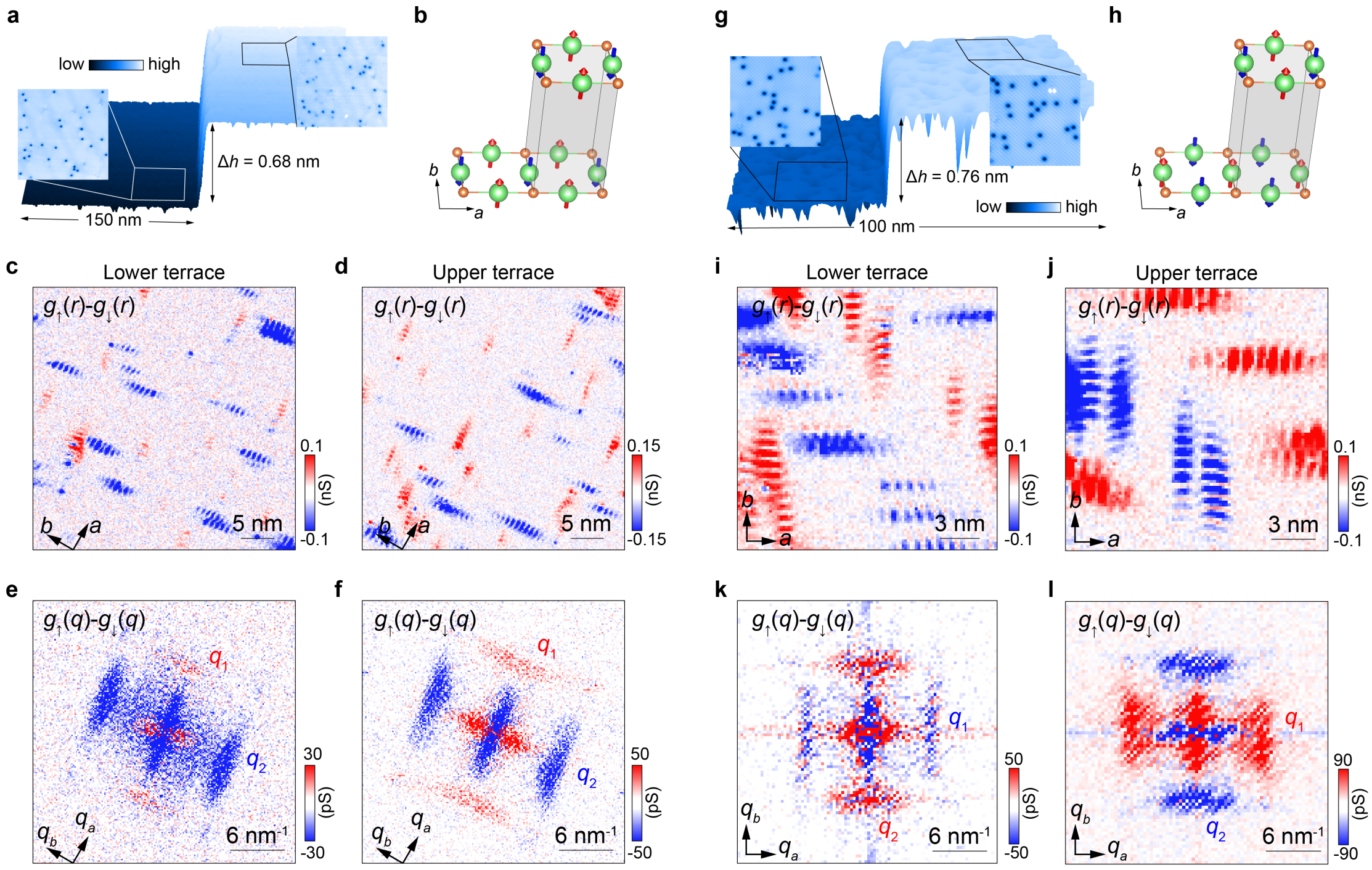

**Fig. 4 | Spin-resolved QPI across unit-cell step edges. a**,**g**, STM topographies ($V_s$ = 1 V, $I_t$ = 30 pA) of a unit-cell step edge separating adjacent $V_2O$ layers. Insets show atomically resolved images acquired on the lower and upper terraces ($V_s$ = 300 mV, $I_t$ = 300 pA). The measured step heights are 0.68 nm (**a**) and 0.76 nm (**g**). **b**,**h**, Schematic C-type (**b**) and G-type (**h**) magnetic structures across a unit-cell step edge. Red and blue arrows denote spin-up and spin-down moments on the V sites, respectively. **c**,**d**, Real-space difference maps $g_{\uparrow}(r) - g_{\downarrow}(r)$ acquired at −6 meV on the lower (**c**) and upper (**d**) terraces shown in **a**. **e**,**f**, Fourier-transformed difference maps $g_{\uparrow}(q) - g_{\downarrow}(q)$ obtained from **c** and **d**, respectively. **i**,**j**, Real-space difference maps $g_{\uparrow}(r) - g_{\downarrow}(r)$ acquired at −6 meV on the lower (**i**) and upper (**j**) terraces shown in **g**. **k**,**l**, Fourier-transformed difference maps $g_{\uparrow}(q) - g_{\downarrow}(q)$ obtained from **i** and **j** terraces, respectively. The corresponding spin-resolved QPI datasets are shown in Extended Data Fig. 3.